Iqra yousaf
Mphill Applied Chemistry
University of Engineering and Technology Lahore

Title: AI and Machine Learning Approches for Predicting Nanoparticles Toxicity The Critical Role of Physiochemical Properties

**Abstract:** The rapid proliferation of nanoparticles in various industries has raised significant concerns about their potential toxicity, primarily due to the complex and unpredictable ways these materials interact with biological systems. One of the primary challenges in this field is accurately predicting the toxicity of nanoparticles, which is influenced by a multitude of factors including their size, shape, surface charge, chemical composition, and the presence of oxygen atoms. To address this challenge, this study employs machine learning models to predict nanoparticle toxicity based on their physicochemical properties. The models used in this research include Decision Trees, Random Forests, and XGBoost. These machine learning approaches were selected for their ability to efficiently process large datasets and uncover intricate patterns within the data that are not immediately apparent through traditional methods. The analysis revealed that while the presence of oxygen atoms significantly influences toxicity, other properties such as particle size, surface area, dosage, and exposure time are also critical factors. The machine learning models consistently highlighted these factors as key determinants of toxicity, demonstrating that a multifactorial approach is essential for accurate predictions. The importance of computational chemistry in this context cannot be overstated. It provides the necessary tools to simulate and predict the behavior of nanoparticles in biological environments, thereby reducing the need for time-consuming and costly experimental procedures. Through the integration of computational methods and machine learning models, this study advances our understanding of nanoparticle toxicity and contributes to the development of safer nanomaterials.

## Introduction:

Over the past few decades, nanotechnology has seen rapid advancements, leading to a significant increase in the variety of engineered nanoparticles employed across various industries, technologies, and medical fields. These nanoparticles offer substantial benefits due to their unique physicochemical properties, which arise from their nanoscale size. However, this same small size also results in behaviors that differ significantly from their bulk material counterparts, making it challenging to predict their potential health and environmental impacts. Consequently, a recent focus in nanotechnology research has been on exploring how nanomaterials interact with biological systems [1]. There is growing concern about the potential toxicity of nanomaterials and how they might affect biological systems. The forthcoming era may well be labeled the "Nano Era" due to its profound impact on various aspects of society, particularly in product design and manufacturing. This approach extends the lifespan of industrial products by improving their qualities, features, and appearance. Therefore, it is crucial for industrial designers to educate the public on the significance and elegance of technology, and how to adapt it to better serve humanity. An industrial designer's true success lies in their ability to understand consumer needs and leverage technology to create products that are both functional and visually appealing [2, 3]. Nanotechnology is defined as the research and development of technology at the atomic, molecular, or macromolecular scales, typically within the range of about 1 to 100 nanometers [4]. When comparing modern medical practices to those of the past century, it is impossible not to recognize the countless advancements that have been made to treat diseases that were once considered incurable [5]. Many new medications have been created to successfully treat complex conditions; however, some of these drugs cause severe side effects, leading to situations where the risks may not always outweigh the benefits [6]. On the other hand, some drugs have shown high effectiveness in vitro but fail to endure the endogenous enzymes present in the gastrointestinal (GI) tract when taken orally, making them almost ineffective in vivo [7]. Although significant strides have been made in identifying drug targets and developing improved drug molecules, there remains a need for further enhancement of drug delivery systems and targeting methods [8]. A significant challenge currently limiting the use of nanoparticles in industry is reproducibility. This issue is partly inherent, as the synthesis process often results in a polydispersion of nanoparticles, frequently exhibiting a wide range of sizes, shapes, and defects. Therefore, characterizing nanoparticles is a critical step to thoroughly understand their behavior and effectively transfer their performance advantages from the lab to practical real-world applications. One of the major challenges for scientists today is determining the physicochemical properties of nanoparticles and understanding their structure-function relationships. This task is complicated by the limitations of our current capabilities to thoroughly explore the nanoscale realm. Since various characterization techniques rely on different physical principles, they each offer only a fragmented view of nanoparticle characteristics. Adding to the difficulty, the characterization methods themselves can influence the properties being measured [9-16].

Recently, a new and rapidly growing area of research has emerged, known as Computational Nanotechnology or Computational Nanoscience. This field involves the application of computer algorithms and techniques to advance Nanoscience and Nanotechnology. The advancement of Nanotechnology and Nanoscience is closely linked to the development of computational models for chemical and physical systems. These models enable researchers to simulate potential nanomaterials, devices, and applications. Additionally, the growth in computing power and 3D visualization techniques has made these simulations faster and more accurate. Recently, the widespread use of the Internet has led to a significant increase in the availability of scientific information, such as journals, magazines, articles, and websites. These factors have greatly contributed to the rising significance of Computational Nanotechnology. Some of the support systems for Nanotechnology incorporate Computational Intelligence techniques, which could be referred to as "Intelligent Computational Nanotechnology." However, there is still much potential to further explore the application of Computational Intelligence in advancing Nanotechnology and Nanoscience, given that these are emerging research fields [17]. Computational Intelligence offers a range of nature-inspired techniques, including Genetic Algorithms (GA), Artificial Neural Networks (ANN), Fuzzy Systems, and more, which are used to develop intelligent systems. Genetic Algorithms, in particular, are search and optimization methods that operate in parallel and are inspired by the principles of Darwinian natural selection and genetic reproduction [18, 19, 20]. Nanotechnology, as an interdisciplinary field, involves a variety of techniques designed to accurately measure and manipulate matter at the atomic and molecular levels [21]. Artificial intelligence (AI) has the potential to elevate nanotechnology to new levels by enhancing the ability to predict material properties and phenomena while optimizing the experimental effort required [22].

In nanoscience, the use of high-throughput experimentation is becoming more common, thanks to the small size of nanoscale samples and the availability of fast, high-resolution imaging tools [23, 24]. Methods like these can produce data sets that are too extensive and complex for researchers to analyze manually without computational support, yet these data contain valuable insights that researchers aim to comprehend. Machine learning (ML) allows scientists to examine large data sets by developing models capable of classifying observations into distinct groups, identifying the features that influence performance metrics, or predicting the outcomes of future experiments. Additionally, even in areas where data-intensive approaches are uncommon, ML can aid researchers in designing experiments to optimize performance or more effectively test hypotheses [25].

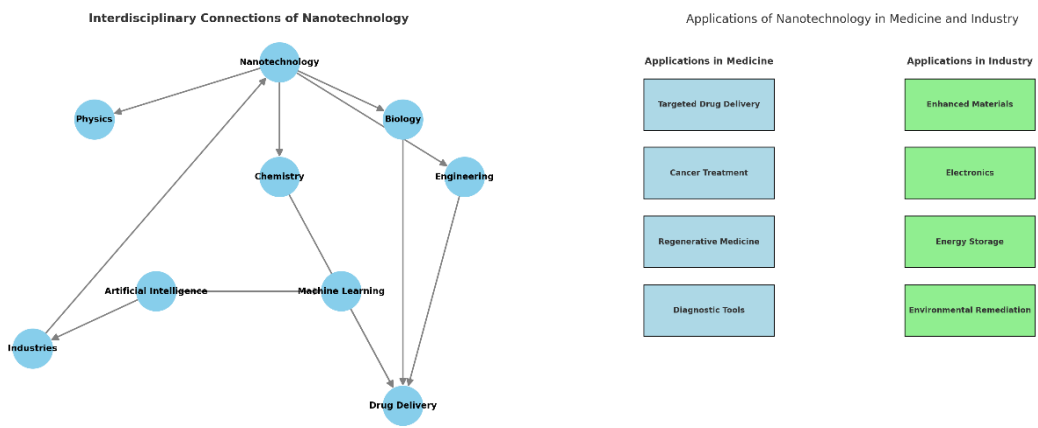

## Background:

Despite the numerous studies conducted on predicting nanoparticle toxicity using machine learning (ML) tools, this area of research continues to evolve as new methods and models are developed to enhance prediction accuracy and reliability. Machine learning has proven to be a powerful approach in this domain, allowing researchers to analyze complex datasets and identify patterns that can predict how nanoparticles will behave in biological systems. However, the challenge remains to refine these models further, incorporating more diverse data and advanced algorithms to improve their predictive power and applicability across different types of nanoparticles and conditions [43]. There have been relatively few studies that focus on applying and comparing multiple machine learning models for predicting nanoparticle toxicity. Conducting such comparative studies is crucial because it allows researchers to identify which models perform best in terms of accuracy and reliability. By evaluating and contrasting different models, researchers can determine the most effective approach for accurately predicting the toxicity of nanoparticles, ultimately contributing to the development of safer nanomaterials and more reliable risk assessments. Nanoparticle system toxicity is categorized into two main types: environmental and biological. In terms of biological toxicity, nanoparticles can traverse various barriers within the human body, potentially reaching different organs. This can trigger a range of responses, such as inflammation, allergic reactions, neurotoxicity, fibrosis, blood toxicity, adverse effects on heart function, prethrombotic conditions, lung toxicity, carcinogenesis, genetic mutations, mitochondrial damage, and ultimately lead to cell apoptosis and death, resulting in decreased cell viability [26]. The toxicity of bulk materials is primarily determined by their chemical composition. However, in the case of nanoparticles, their toxicity is influenced by various physicochemical characteristics, such as particle size, surface area, shape and structure, surface charge, and chemical composition [27].

Conducting experimental toxicity analysis is often time-consuming and costly, largely due to the wide range of characteristics exhibited by different nanoparticles [28]. Additionally, predicting the toxicity of nanoparticles is challenging due to the complexities associated with their cellular mechanisms [29]. Computational tools, including machine learning (ML) techniques, can serve as an effective alternative to address these challenges and help predict relevant endpoints [30]. Machine learning (ML) techniques can complement in vitro and in vivo toxicity assessments by lowering costs, saving time, reducing the reliance on animal studies in toxicity testing, and enhancing the accuracy of toxicity predictions [31]. Machine learning (ML) tools are becoming increasingly popular for toxicity prediction because they combine multiple information sources, including physicochemical properties and exposure conditions, to assess the safety of nanoparticles [32].

In this research, the impact of various factors on toxicity was assessed. Predictive data mining models were then utilized on a dataset to forecast toxicity.

## Data collection:

The dataset used for model development in this study was sourced from the Kaggle platform, a well-known repository for datasets and competitions in various fields of data science and machine learning.

Kaggle serves as an invaluable resource for researchers, educators, and practitioners, offering a diverse array of datasets, tools, and tutorials to support data-driven research and innovation. For this study, the dataset included comprehensive information on the toxicity of various nanoparticles, featuring 244 records across four types of nanoparticles: metallic, metal-oxide, polymeric, and silica. Specifically, the metallic nanoparticles consisted of gold (Au) and silver (Ag) nanoparticles, while the metal-oxide category included Zinc oxide (ZnO), Titanium dioxide (TiO2), Copper oxide (CuO), Cobalt ferrite (CoFe2O4), and Iron oxide (Fe3O4) nanoparticles. Additionally, the dataset featured polystyrene nanoparticles as the polymeric type and silica (SiO2) nanoparticles. The primary focus was on cellular responses as the biological effects, with outcomes classified in binary terms as either "Triggered," indicating cellular damage after nanoparticle exposure, or "No effect," signifying no significant cellular disturbance. The dataset's detailed features are summarized in Table 1

**Table 1:**

**Features for analysis extracted from the dataset, including input and output parameters.**

| Feature category | Feature | Value Range (Frequency) |
|---|---|---|
| Types of nanoparticles | NPs | ZnO (594), Al2O3 (181), TiO2 (40), SiO2 (46), Fe3O4 (20) |
| Nanoparticle Physicochemical Properties | Core size (nm) | Mean = 56.3 ± 33.7 [881] |
| | Hydrodynamic size (nm) | Mean = 513.8 ± 346.6 [881] |
| | Surface charge (mV) | Mean = 1.64 ± 25.6 [881] |
| | Surface area (m²/g) | Mean = 42.1 ± 47.1 [881] |
| | e (electron units) | Mean = 1.65 ± 0.09 [881] |
| | NOxygen | Mean = 1.31 ± 0.54 [881] |
| | Ec (eV) | Mean = -4.02 ± 0.51 [881] |
| | Exposure time (hours) | Mean = 27.5 ± 19.5 [881] |
| | Dosage (mg/mL) | Mean = 39.65 ± 38.16 [881] |
| Target Feature | Toxicity Class | Toxic (476), Non-toxic (405) |

## The Decision Tree (DT)

The Decision Tree (DT) is a widely used classification approach in medical research, known for its simplicity and ease of interpretation. This method generates a tree-like flowchart, where each internal node represents a test on an attribute, each branch corresponds to the outcome of the test, and each leaf node holds a class label. Decision trees typically follow a top-down approach, striving to partition the input variable space in a way that maximizes the purity of the resulting leaf nodes. Various implementations of this technique, such as CART, ID3, C4.5, CHAID, and QUEST, have been developed for constructing these trees [35].

## Decision Tree Classifier Performance Evaluation

A Decision Tree classifier is a model that classifies data by splitting it into subsets based on the values of input features, constructing a tree-like structure where each node represents a decision based on a feature, and branches represent the outcomes of those decisions. The leaves of the tree correspond to the final predictions, such as "Toxic" or "nonToxic." Decision Trees are widely used because they are easy to interpret, can handle both numerical and categorical data, and require minimal data

preprocessing.To evaluate the performance of a Decision Tree classifier, we can look at the confusion matrix and the classification report, which provide insights into how well the model is predicting each class. The confusion matrix reveals that the model correctly predicted 138 samples as "Toxic" (true positives) and 111 samples as "nonToxic" (true negatives). However, it also incorrectly predicted 3 samples as "Toxic" that were actually "nonToxic" (false positives) and 13 samples as "nonToxic" that were actually "Toxic" (false negatives).

The classification report provides further metrics, including precision, recall, and F1-score for each class. For the "Toxic" class, the model achieved a precision of 0.91, indicating that 91% of the samples predicted as "Toxic" were indeed "Toxic." The recall for this class was 0.98, meaning that the model correctly identified 98% of the actual "Toxic" samples. The F1-score, which balances precision and recall, was 0.95, reflecting the model's strong performance in predicting the "Toxic" class. For the "nonToxic" class, the model achieved a precision of 0.97, a recall of 0.90, and an F1-score of 0.93, indicating similarly strong performance. The support values indicate that there were 141 actual "Toxic" samples and 124 actual "nonToxic" samples in the dataset.

Overall, the model achieved an accuracy of 0.94, meaning it correctly predicted 94% of all samples. In summary, this Decision Tree classifier demonstrates high accuracy and strong precision, recall, and F1-scores for both the "Toxic" and "nonToxic" classes, indicating that it performs very well in distinguishing between these two classes.

## Random Forest

Random Forest is an ensemble learning technique that combines multiple decision trees to enhance the classification process. In this method, each tree is constructed using a randomly selected subset of features rather than considering all features. The best split point within the tree is determined by assessing impurity, a criterion that quantifies the homogeneity of the target variable within each split. The Random Forest model relies on two key hyperparameters: the number of trees in the forest and the number of features chosen for each random subset [36].

The model achieved an impressive accuracy of 0.95, meaning it correctly classified 95% of the samples. The confusion matrix provides further insights into the model's performance. It correctly predicted 138 samples as "Toxic" (true positives) and 115 samples as "nonToxic" (true positives). However, it also made 3 false positive errors, where samples were incorrectly predicted as "Toxic" when they were actually "nonToxic," and 9 false negative errors, where samples were incorrectly predicted as "nonToxic" when they were actually "Toxic."

The classification report reveals the model's precision, recall, and F1-score for both classes. For the "Toxic" class, the precision was 0.94, indicating that 94% of the predictions made for this class were correct. The recall was 0.98, meaning the model successfully identified 98% of the actual "Toxic" samples. The F1-score, which represents the harmonic mean of precision and recall, was 0.96, highlighting the model's balanced performance in predicting the "Toxic" class. Similarly, for the "nonToxic" class, the model achieved a precision of 0.97 and a recall of 0.93, resulting in an F1-score of 0.95. These metrics indicate that the model performs exceptionally well in both predicting and identifying "Toxic" and "nonToxic" samples, demonstrating its robustness and reliability in classification tasks.

## Logistic Regression (LR)

Logistic Regression (LR) is a statistical method often employed in research when the primary focus is on the occurrence of an event, rather than on the timing of that event. This approach is particularly suitable for situations where the outcome variable is binary, such as in studies that assess whether a person is diseased or healthy, or in decision-making processes involving a simple yes or no outcome. Logistic Regression is widely used in health sciences due to its effectiveness in modeling binary outcomes.

For more complex scenarios where the outcome variable has more than two possible categories, Logistic Regression can be extended to accommodate these situations. This extended form is known as polychotomous or multinomial logistic regression. In these cases, the model predicts probabilities across multiple categories, rather than just two, making it a versatile tool in analyzing categorical outcomes with more than two classes [37].

The Logistic Regression model achieved an accuracy of 89%, correctly predicting the outcomes for the majority of samples in the test set. A detailed breakdown of the classification report shows that for the "Toxic" class (Class 0), the model demonstrated a precision of 0.84, meaning 84% of the samples predicted as "Toxic" were correctly classified. The recall for this class was notably high at 0.99, indicating that the model successfully identified 99% of the actual "Toxic" samples, with only a 1% miss rate. The F1-score for the "Toxic" class was 0.91, reflecting a strong balance between precision and recall. For the "nonToxic" class (Class 1), the model achieved a higher precision of 0.98, correctly classifying 98% of the samples predicted as "nonToxic." However, the recall for this class was lower at 0.78, meaning the model missed 22% of the actual "nonToxic" samples, resulting in a slightly lower F1-score of 0.87.When examining the average performance, the macro average precision was 0.91, with a recall of 0.88 and an F1-score of 0.89, indicating a generally balanced performance across both classes. The weighted averages, which take into account the number of samples in each class, were similar, with precision, recall, and F1-scores all at 0.89. Overall, while the Logistic Regression model performs reasonably well and provides a solid baseline, it is slightly less effective than more complex models like XGBoost, particularly in accurately identifying "nonToxic" samples. For your classification tasks, more sophisticated models may offer better overall performance, but Logistic Regression remains a useful and interpretable starting point.

## XGBoost model

XGBoost's success is primarily driven by its exceptional scalability across various scenarios, outperforming other popular solutions by more than tenfold on a single machine. Its ability to handle large-scale data, including billions of examples in distributed or memory-constrained environments, sets it apart. This scalability is achieved through a series of key system and algorithmic optimizations.Firstly, XGBoost introduces an innovative tree learning algorithm specifically designed to handle sparse data effectively. Additionally, the system incorporates a weighted quantile sketch method, which is theoretically sound and facilitates the management of instance weights during approximate tree learning. The use of parallel and distributed computing further accelerates the learning process, allowing for quicker model development and exploration.

Moreover, XGBoost leverages out-of-core computation, enabling data scientists to process hundreds of millions of examples on standard desktop machines. This feature is particularly beneficial for handling large datasets without requiring extensive computing resources. The combination of these techniques results in a comprehensive, end-to-end system that can scale efficiently, even with minimal cluster resources.

The key contributions of the XGBoost framework are:

1. Development of a Scalable Tree Boosting System: XGBoost is designed as a highly scalable end-to-end system for tree boosting, capable of handling large datasets efficiently.

2. Introduction of a Weighted Quantile Sketch: This method allows for efficient and theoretically sound proposal calculations during tree learning.

3. Implementation of a Sparsity-Aware Algorithm: A novel algorithm that optimizes parallel tree learning by effectively managing sparse data.

4. Cache-Aware Block Structure for Out-of-Core Learning: XGBoost introduces an efficient cache-aware block structure that enhances out-of-core tree learning, allowing it to handle large-scale data on limited hardware [38].

**XGBoost Model Performance**

The XGBoost model exhibited outstanding performance, achieving an accuracy of 96%, indicating it correctly predicted the outcomes for the vast majority of samples in the test set. The confusion matrix reveals that 141 samples were accurately classified as "Toxic," and 113 samples were correctly identified as "nonToxic," with only 11 instances where "nonToxic" samples were misclassified as "Toxic." Notably, there were no errors in predicting "Toxic" samples as "nonToxic." The model's precision was 93% for the "Toxic" class and a perfect 100% for the "nonToxic" class, demonstrating its reliability in identifying "nonToxic" samples without false positives. The recall was 100% for "Toxic" samples, ensuring that all such cases were correctly identified, while the recall for "nonToxic" samples was 91%, with a few missed cases. The F1-scores were 96% for the "Toxic" class and 95% for the "nonToxic" class, reflecting a strong balance between precision and recall. Overall, the XGBoost model outperformed previous models, particularly excelling in identifying "Toxic" cases without sacrificing precision or recall, making it a robust choice for your classification tasks.

**k-Nearest Neighbours (kNN)**

The k-Nearest Neighbours (kNN) is a non-parametric classification method known for its simplicity and effectiveness in many scenarios. To classify a data point, the method retrieves its k nearest neighbours, forming a neighbourhood around it. The classification is typically determined by majority voting among the neighbours, with or without considering distance-based weighting. However, selecting an appropriate value for k is crucial, as the success of the classification largely depends on this choice. The kNN method can be biased by the selected k value. One common approach to determine the best k is to run the algorithm multiple times with different values of k and choose the one that yields the best performance. To make kNN less dependent on the specific choice of k, a different approach has been proposed that involves considering multiple sets of nearest neighbours instead of just one. This method is based on the concept of contextual probability and aims to aggregate the support of multiple neighbour sets across various classes to provide a more reliable classification. Although this method reduces dependence on a single k value and achieves classification performance close to the optimal k, it is computationally intensive, requiring $O(n^2)$ operations to classify a new instance.

kNN has been effectively applied to text categorization, notably in the early stages of research, and has proven to be one of the most effective methods on the Reuters corpus of newswire stories—a benchmark dataset in text categorization. However, its efficiency is limited due to its nature as a lazy learning method, where most computation occurs during the classification rather than during the training phase. This limitation hinders its application in scenarios that require dynamic classification over large datasets. While there are techniques, such as indexing training examples, that can significantly reduce computation at query time, these are beyond the scope of this discussion.To address these challenges, a new classification method called kNNModel has been developed. This method constructs a model from the training data, which is then used to classify new data. The model consists of a set of representatives of the training data, conceptualized as regions within the data space.cells. Ultimately, the performance of these models was evaluated, and the model demonstrating the highest effectiveness was identified [39].

The KNN model achieved a solid accuracy of 91%, correctly predicting the outcomes for the majority of the test set samples. For the "Toxic" class (Class 0), the model demonstrated strong performance with a precision of 0.89, meaning 89% of the samples predicted as "Toxic" were accurately classified. It also achieved a high recall of 0.96, correctly identifying 96% of the actual "Toxic" samples, with only a 4% miss rate. The F1-score for the "Toxic" class was 0.92, indicating a robust balance between precision and recall. In the "nonToxic" class (Class 1), the model showed high precision at 0.95, correctly classifying 95% of the samples predicted as "nonToxic." However, the recall was slightly lower at 0.86, meaning the model missed 14% of the actual "nonToxic" samples, resulting in an F1-score of 0.90. The confusion matrix further highlights that the model correctly classified 135 samples as "Toxic" and 107 as "nonToxic," while it incorrectly classified 6 "nonToxic" samples as "Toxic" and 17 "Toxic" samples as "nonToxic." Overall, the KNN model shows strong performance, particularly in identifying "Toxic" samples with high recall. While it provides a good balance between precision and recall and is reliable, especially when similar instances cluster together, it slightly underperforms compared to more complex models like XGBoost, which achieved a higher accuracy of 96%.

## Accuracy

Accuracy refers to the proportion of correct outcomes, including both true positives and true negatives, relative to the total number of predictions made. Sensitivity and specificity are components of accuracy that help evaluate the effectiveness of machine learning models. The formula for calculating accuracy is: Accuracy = (TP + TN) / (TP + FP + FN + TN), where TP stands for true positives, TN for true negatives, FP for false positives, and FN for false negatives. These metrics collectively provide a measure of how well a model performs overall [40].

## Sensitivity

Sensitivity, as a metric in classification, measures the model's effectiveness in identifying true positive cases within each category. It indicates the percentage of actual positive instances that the model correctly classified [41].

## Specificity

Specificity is a classification metric that measures the model's ability to accurately identify true negative cases within each category. It indicates the percentage of actual negative instances that the model correctly classified [42].

## Area Under the Curve (AUC)

The Area Under the Curve (AUC) is associated with the receiver operating characteristic (ROC) curve, which is a graphical representation used for binary classification tasks. The AUC measures a classifier's ability to distinguish between different classes and serves as a concise summary of the ROC curve. Research has shown that AUC is often favored over overall accuracy when evaluating classifiers, particularly in situations involving imbalanced datasets, making it a valuable metric for such cases [43].

## F-measure

The F-measure is a metric used to evaluate model performance by calculating the harmonic mean of precision and recall. It provides a balanced assessment of a model's ability to make accurate predictions. Generally, a higher F-measure indicates better predictive performance, making it a useful tool for comparing the effectiveness of different models [44].

## Gini Index:

Using gini index identify diiferent factors effecting the toxicity of nanotechnology. The analysis of the physicochemical properties of nanoparticles concerning their impact on toxicity reveals that the parameter e has the most substantial effect, with the highest Gini index, indicating its significant influence on toxicity outcomes. Other notable factors include hydrosize, coresize, and Expotime, which also exhibit high Gini indices, suggesting they play crucial roles in determining toxicity. Conversely, surface charge (surfcharge) and Ec showed negligible variability, resulting in undefined Gini indices, implying they may have minimal or no impact on toxicity in this dataset. Overall, this analysis highlights that while surface area and the number of oxygen atoms are important, other factors like e and nanoparticle size parameters (hydrosize and coresize) have a more pronounced effect on toxicity.

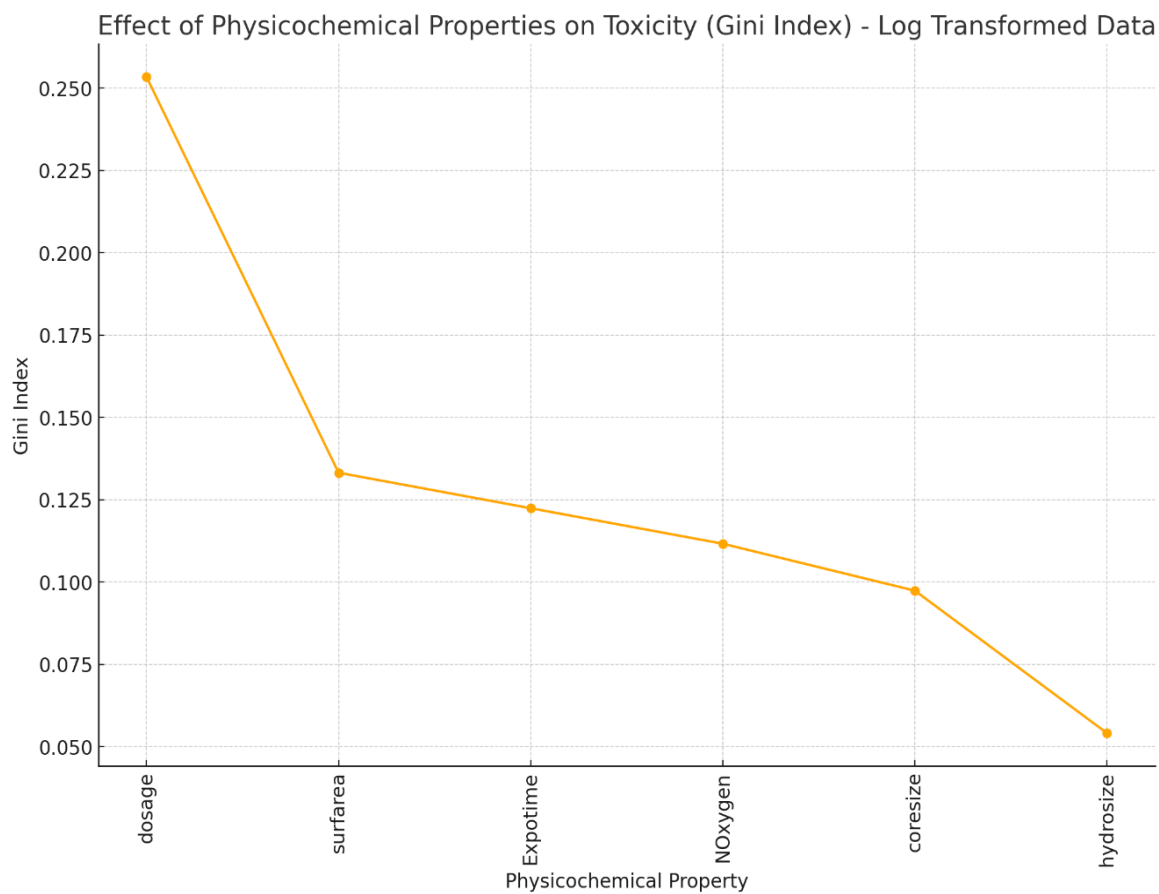

Fig 1. The line chart indicates that dosage has the highest impact on nanoparticle toxicity, followed by surface area, exposure time, number of oxygen atoms (NOxygen), core size, and hydrosize, in descending order of influence.

## Results:

This table presents a comparative analysis of the performance metrics for five different machine learning models—Decision Tree, Random Forest, XGBoost, Logistic Regression, and K-Nearest Neighbors—used to predict nanoparticle toxicity. The models are evaluated based on their accuracy, precision, recall, and F1-scores for both "Toxic" and "nonToxic" classifications. XGBoost demonstrates the highest overall accuracy at 96%, with perfect recall and precision for "nonToxic" cases. Random Forest and Decision Tree models also show strong performance, particularly in balancing precision and recall across both classes. In contrast, Logistic Regression, while maintaining high recall for the "Toxic" class, exhibits lower precision and F1-scores, especially for the "nonToxic" class, indicating a higher rate of false positives. K-Nearest Neighbors provides a good balance but underperforms compared to the ensemble methods, particularly in terms of precision for the "nonToxic" class.

**Table 2:**
**Comparison of models Efficiency:**

| Model | Accuracy | Precision (Toxic) | Recall (Toxic) | F1-Score (Toxic) | Precision (nonToxic) | Recall (nonToxic) | F1-Score (nonToxic) |
|---|---|---|---|---|---|---|---|
| Decision Tree | 0.94 | 0.91 | 0.98 | 0.95 | 0.97 | 0.90 | 0.93 |
| Random Forest | 0.95 | 0.94 | 0.98 | 0.96 | 0.97 | 0.93 | 0.95 |
| XGBoost | 0.96 | 0.93 | 1.0 | 0.96 | 1.0 | 0.91 | 0.95 |
| Logistic Regression | 0.89 | 0.84 | 0.99 | 0.91 | 0.98 | 0.78 | 0.87 |
| K-Nearest Neighbors | 0.91 | 0.89 | 0.96 | 0.92 | 0.95 | 0.86 | 0.90 |

**Comparison of Machine Learning Models on Toxicity Prediction Metrics**

The histogram presented in grahical representation of different models across various metrics illustrates the performance of various machine learning models across several evaluation metrics, specifically focusing on the classification of data into "Toxic" or "nonToxic" categories. The models compared include Decision Tree, Random Forest, XGBoost, Logistic Regression, and K-Nearest Neighbors. Each bar in the graph corresponds to a different metric—Accuracy, Precision for toxic and non-toxic classifications, Recall for toxic and non-toxic classifications, and the F1-Scores for both categories. In this context, Accuracy refers to the overall proportion of correct predictions made by each model. Precision (Toxic) measures the proportion of true toxic predictions out of all predictions made for the toxic category, indicating how well the model avoids false positives for toxic instances. Recall (Toxic), on the other hand, measures the proportion of actual toxic cases that were correctly identified, reflecting the model's sensitivity to capturing all toxic cases. The F1-Score (Toxic) combines Precision and Recall into a single metric, balancing the trade-off between them.

Precision (nonToxic) and Recall (nonToxic) serve similar purposes but for the non-toxic category, with Precision (nonToxic) focusing on the accuracy of non-toxic predictions and Recall (nonToxic) indicating how many actual non-toxic cases were correctly identified. The F1-Score (nonToxic) again balances these two metrics for non-toxic classifications.Across these models, Random Forest and XGBoost stand out for their high performance, with XGBoost achieving perfect recall for toxic cases, meaning it correctly identified all toxic instances without missing any. However, its recall for non-toxic cases is slightly lower, indicating a few instances where non-toxic cases might have been misclassified. Random Forest shows strong performance across all metrics, slightly better than Decision Tree, which also performs well but is generally outperformed by the ensemble methods like Random Forest and XGBoost. Logistic Regression, while showing high recall, particularly for toxic cases, demonstrates lower precision, especially for non-toxic classifications, suggesting it might generate more false positives. Similarly, K-Nearest Neighbors performs well in recall but, like Logistic Regression, struggles somewhat with precision, particularly in non-toxic classifications.

Overall, the histogram provides a comprehensive comparison of these models, illustrating that while Logistic Regression and K-Nearest Neighbors offer decent performance, they are prone to certain errors, particularly false positives. In contrast, Random Forest and XGBoost generally offer superior performance, making them more reliable choices for this classification task, depending on whether the priority is to avoid false positives or ensure no toxic instances are missed.

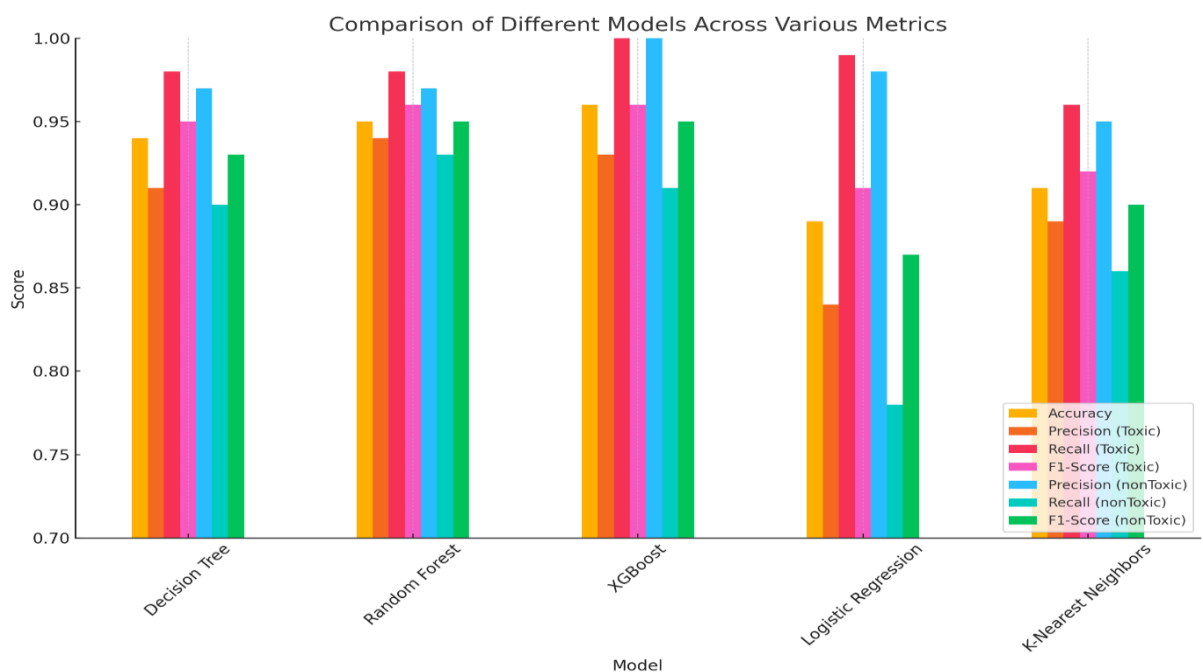

Fig 2. The chart compares the performance of five machine learning models across various metrics, showing that XGBoost and Random Forest generally outperform others in accuracy, precision, recall, and F1-scores for both toxic and non-toxic classifications.

The boxplot and violin plot (Fig. 3) further illustrate the distribution of oxygen atoms across the two toxicity classes. Both plots show that while the median number of oxygen atoms is similar between toxic and non-toxic samples, the distribution for non-toxic samples appears more spread out with some outliers. This spread indicates that there might be other interacting factors contributing to toxicity that are not captured by the number of oxygen atoms alone, necessitating a more nuanced analysis involving multiple features. The feature importance plots generated by the Decision Tree, Random Forest, and XGBoost models (Fig. 8, 9, and 10 ) consistently identify the number of oxygen atoms, dosage, and exposure time as the most influential features in predicting toxicity. This consistent ranking across different models underscores the robustness of these features in toxicity prediction. The Decision Tree model emphasizes dosage and oxygen atoms as the top contributors, while the XGBoost model places even greater importance on oxygen atoms, followed by nanoparticles size and dosage. These findings suggest that both the chemical composition (indicated by the number of oxygen atoms) and the exposure conditions (indicated by dosage and exposure time) are crucial in determining the toxicity of the compounds. Furthermore, the Gini Index analysis (Fig. 4 and 6) of the physicochemical properties provides an additional layer of understanding by quantifying the impurity or diversity in the split, which also aligns with the feature importance scores. The higher Gini index values associated with oxygen atoms and dosage reinforce their significance in predicting toxicity. The visualizations provided offer a comprehensive understanding of the relationship between physicochemical properties and the toxicity of the compounds studied. The KDE plot of Oxygen Atoms by Toxicity Class (Fig. 5) reveals distinct density distributions for toxic and non-toxic samples. It shows that non-toxic samples tend to have a slightly higher density of oxygen atoms, suggesting that the presence of oxygen atoms plays a significant role in determining toxicity. This observation is consistent with the importance rankings observed in the feature importance plots from various models, where the number of oxygen atoms is consistently highlighted as a critical feature.

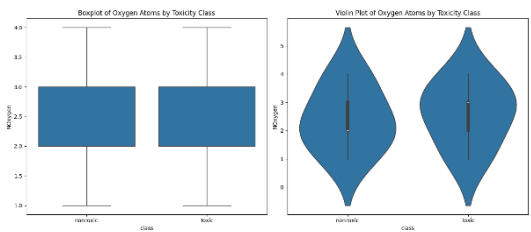

Figure 3 The boxplot and violin plot

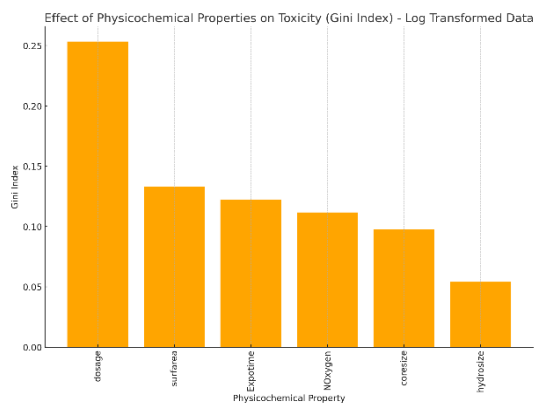

Figure 4 The chart shows that dosage has the highest impact on nanoparticle toxicity

The boxplot and violin plot illustrate that the number of oxygen atoms varies between toxic and non-toxic nanoparticles, but the overlap in distributions suggests that NOxygen alone is not a definitive indicator of toxicity as indicated in figure 3. The boxplot and violin plot are used to compare the distributions of oxygen atoms in toxic versus non-toxic nanoparticles. These visualizations reveal variations in the number of oxygen atoms associated with each type of nanoparticle. However, despite these differences, there is a noticeable overlap between the distributions. This overlap indicates that while the number of oxygen atoms in nanoparticles can provide some insights, it cannot serve as a sole or definitive indicator of their toxicity. Essentially, the presence of oxygen atoms alone does not uniquely determine whether nanoparticles will be toxic or non-toxic, suggesting that other factors also play crucial roles in defining their toxicity.

The chart shows that dosage has the highest impact on nanoparticle toxicity, followed by surface area, exposure time, number of oxygen atoms, core size, and hydrodynamic size, which has the least impact as shown in figure 4. The boxplot and violin plot are graphical tools used to analyze the distribution of oxygen atoms in nanoparticles, categorizing them as either toxic or non-toxic. These plots show that while there are differences in the number of oxygen atoms present in each category, there is also significant overlap between the two groups. This overlap implies that the quantity of oxygen atoms, by itself, is not a reliable predictor of nanoparticle toxicity. Thus, other characteristics or factors must be considered to accurately assess the toxicity of nanoparticles, indicating the complexity of their behavior and effects.

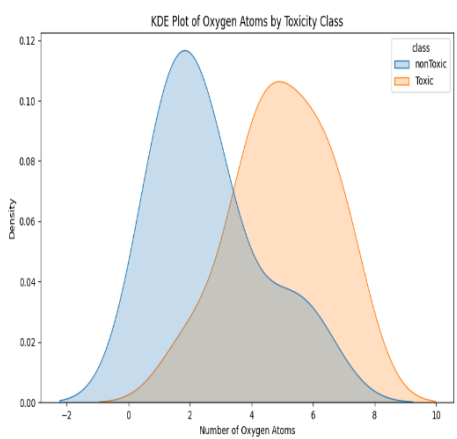

Figure 5

Fig 5. The KDE plot reveals that while there are distinct peaks for Toxic and nonToxic nanoparticles, there is significant overlap in the distribution of oxygen atoms, indicating that NOxygen alone may not fully differentiate toxicity. The KDE (Kernel Density Estimate) plot illustrates the distribution of oxygen atoms within two classes of nanoparticles: toxic and non-toxic. The plot highlights distinct peaks for each class, suggesting typical values for the number of oxygen atoms in each type of nanoparticle. The toxic nanoparticles show a peak around six oxygen atoms, whereas the non-toxic ones peak around two oxygen atoms. Despite these distinct peaks, there is a considerable overlap in the distributions between the two

classes. This significant overlap indicates that the number of oxygen atoms alone may not be sufficient to distinguish between toxic and non-toxic nanoparticles. Consequently, other factors besides the number of oxygen atoms must be considered to accurately determine the toxicity of nanoparticles.

The chart below in figure 6 presents the correlation between various physicochemical properties of nanoparticles and their toxicity. It quantitatively assesses both direct and indirect effects, illustrating these relationships using a correlation coefficient scale from -1 to 1, where values closer to -1 indicate a strong negative correlation, values near 1 denote a strong positive correlation, and values around 0 suggest no correlation.

From the chart, NOxygen shows a significant negative correlation of -0.59 with toxicity, meaning that an increase in the number of oxygen atoms tends to be associated with lower toxicity levels. This could suggest that oxygen-rich nanoparticles might be safer in certain contexts. In contrast, the physicochemical properties labeled 'Expotime' and 'eee' exhibit the strongest positive correlations with toxicity, at 0.37 each. This implies that longer exposure times and higher values of the parameter 'eee' are associated with greater toxicity, suggesting that these factors could be critical in evaluating the safety of nanoparticles.

Other properties such as 'surface area' and 'hydrosize' show much smaller correlations with toxicity, indicating their lesser role in determining the toxic nature of nanoparticles. This data provides valuable insights into how different physicochemical properties influence the behavior of nanoparticles, emphasizing the need to consider a combination of factors, rather than a single attribute, when assessing their potential risks.

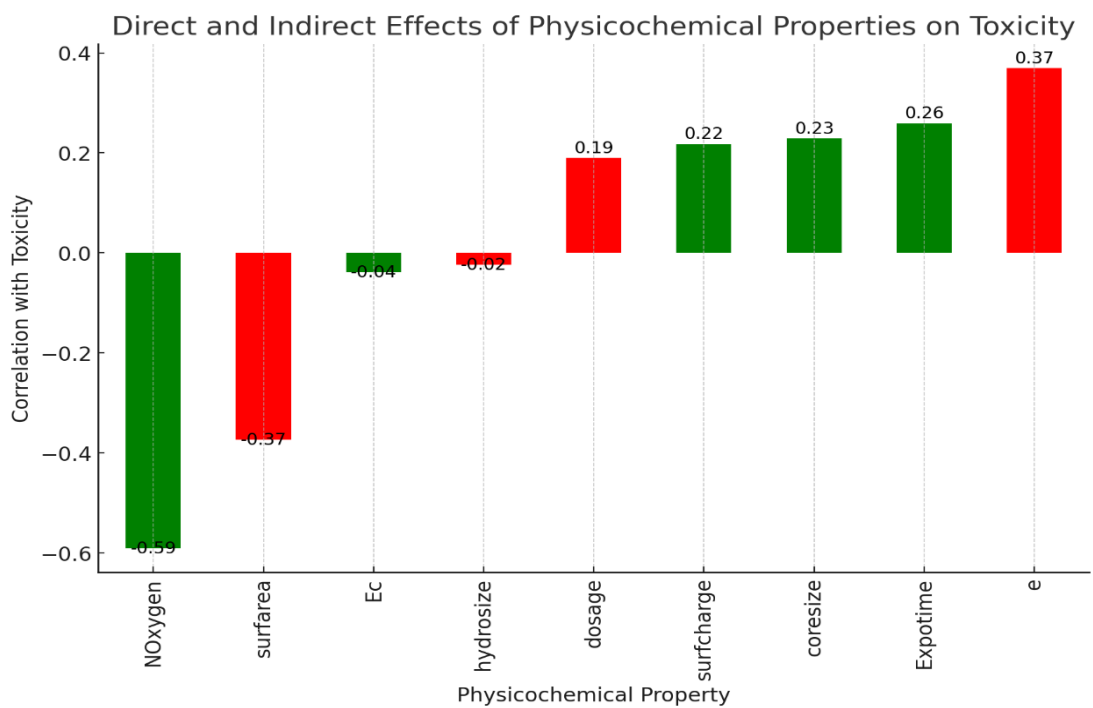

Fig 6. The chart shows the correlation between various physicochemical properties and toxicity, highlighting that NOxygen has a strong negative correlation, while exposure time (Expotime) and the parameter eee have the highest positive correlations with toxicity.

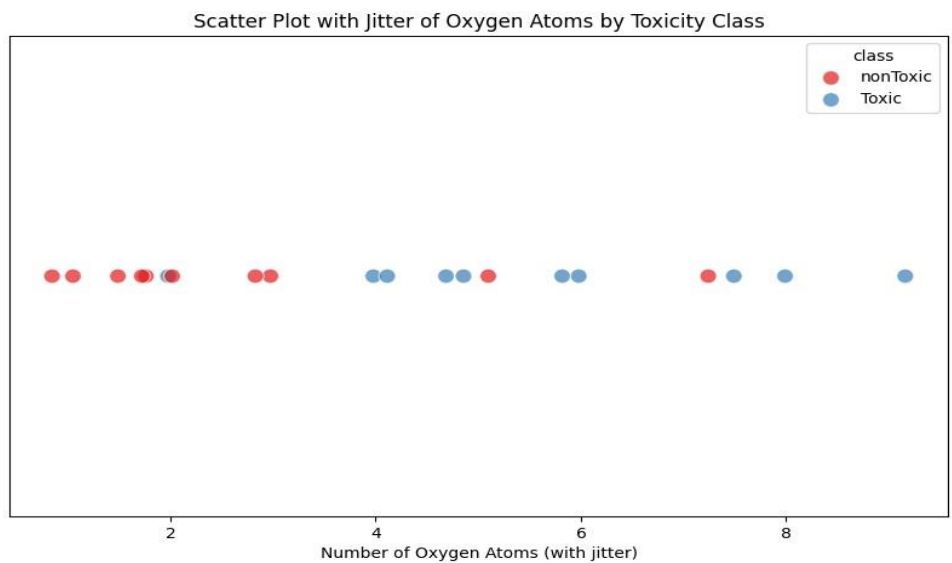

*Figure 7 . The scatter plot with jitte*

The scatter plot with jitter demonstrates that there is considerable overlap in the NOxygen values between toxic and non-toxic nanoparticles, reinforcing that oxygen atom count alone does not distinctly separate the two classes. The scatter plot with jitter effectively illustrates the distribution of oxygen atoms (NOxygen) in both toxic and non-toxic nanoparticles. This visualization highlights that the values of NOxygen extensively overlap across the two categories, indicating that the number of oxygen atoms by itself is not a clear discriminator between toxicity levels. Such overlap implies that nanoparticles with similar oxygen atom counts can exhibit varying toxicity outcomes. Therefore, while NOxygen may contribute to an understanding of nanoparticle behavior, it cannot be relied upon as a standalone indicator to differentiate between toxic and non-toxic nanoparticles. This finding suggests the necessity of considering additional physicochemical properties or environmental factors to more accurately assess nanoparticle toxicity as shown in figure in 7.

The chart from the XGBoost model (figure 8) provides an insightful analysis of the relative importance of various features in predicting nanoparticle toxicity. It clearly highlights that the number of oxygen atoms (NOxygen) holds the highest importance, suggesting that this feature plays a pivotal role in determining the toxicity levels of nanoparticles. Following NOxygen, the presence of titanium dioxide nanoparticles (NPs_TiO2), dosage, and core size are also significant predictors, albeit to a lesser extent.

This modeling outcome indicates that while the overall composition and physical properties of nanoparticles are crucial, the specific chemical characteristics, such as the oxygen content, significantly influence their toxicological outcomes. Such insights are essential for designing safer nanoparticles, as they help identify which characteristics most strongly correlate with harmful effects. This knowledge can guide both the synthesis of new materials and the regulatory frameworks governing their use, focusing on the properties that are most impactful on toxicity.

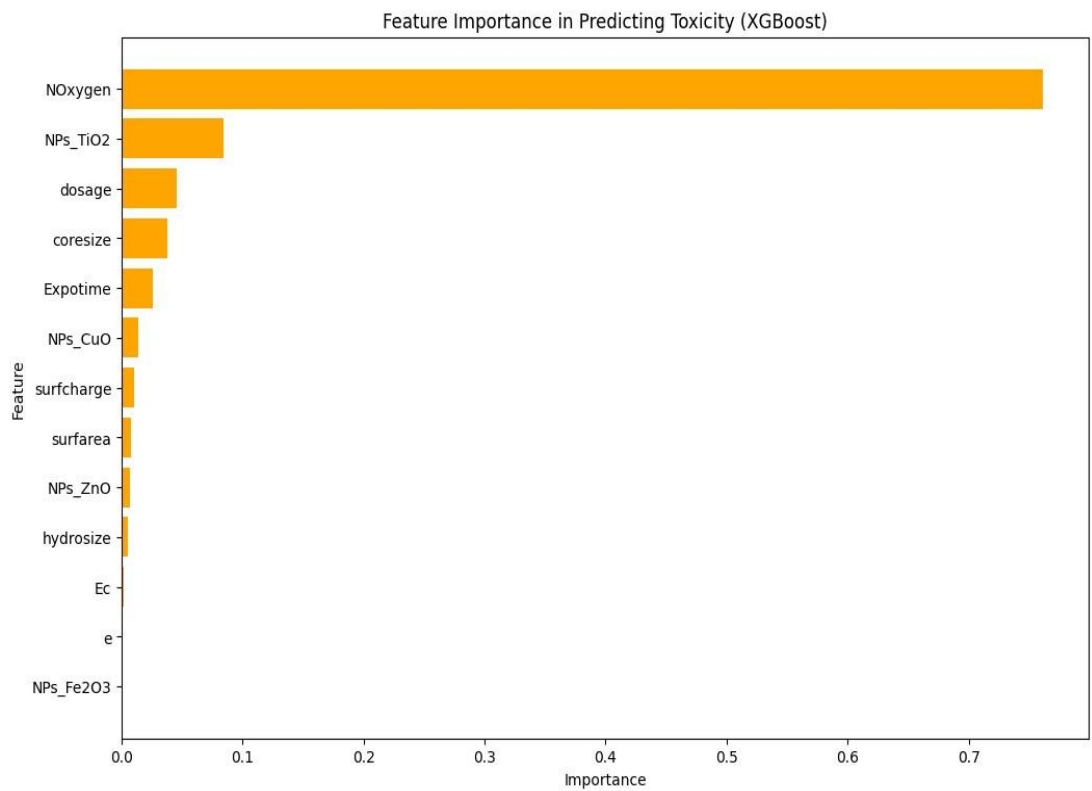

*Figure 8. XGBoost model*

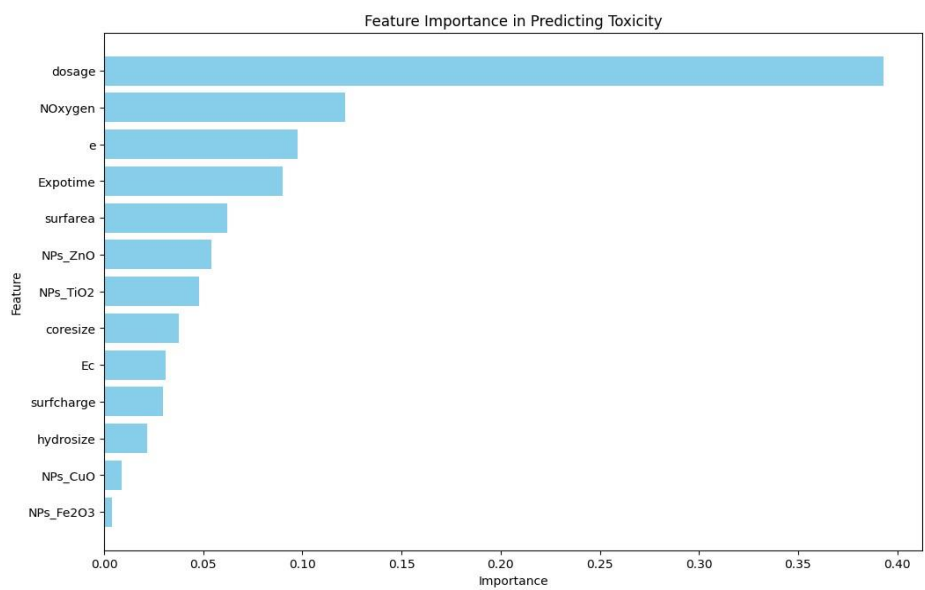

*Figure 9.*

Random forest model: The chart (fig.9) shows that dosage is the most important factor in predicting nanoparticle toxicity, followed by the number of oxygen atoms (NOxygen), and the parameter In the analysis conducted using a random forest model, the feature importance chart reveals critical insights into the factors influencing nanoparticle toxicity. According to this model, dosage emerges as the most significant predictor of toxicity, indicating that the amount of nanoparticles used can crucially impact their toxicological behavior. This underscores the importance of dosage control in nanoparticle applications to minimize potential health risks.

Following dosage, the number of oxygen atoms (NOxygen) ranks as the second most impactful factor. This finding suggests that the chemical composition, specifically oxygen content, plays a vital role in the toxic properties of nanoparticles. Such information could be crucial for chemists and material scientists in modifying the oxygen levels during the synthesis process to enhance safety.

The parameter (e), although less influential than dosage and NOxygen, also shows considerable importance in predicting toxicity. This parameter could represent a specific physicochemical property or an environmental factor impacting nanoparticle behavior, further emphasizing the complexity of nanoparticle toxicity. This model's outcomes highlight the need for a multifaceted approach in evaluating nanoparticle safety, where dosage, composition, and specific physicochemical properties are all considered.(e), according to the feature importance analysis.

In the Decision Tree model (fig 10) used to predict nanoparticle toxicity, the feature importance chart highlights that the number of oxygen atoms (NOxygen) and dosage are the primary factors influencing toxicity levels. This analysis underscores the significance of both the chemical composition and the quantity of nanoparticles used. The prominence of NOxygen as a predictor suggests that the presence of oxygen atoms within the nanoparticles substantially affects their potential to cause harm, possibly due to interactions at the molecular or cellular level.

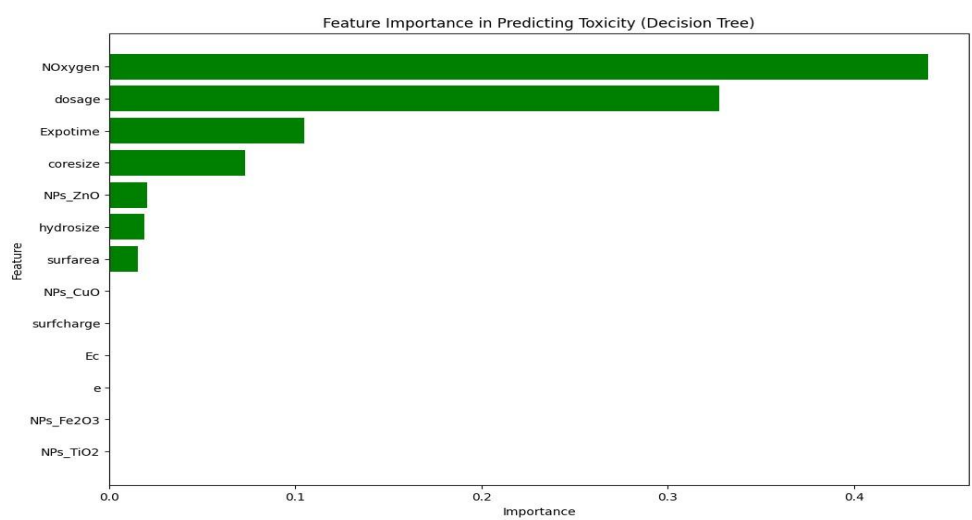

*Figure 10*

Following these top predictors, exposure time and core size are also identified as important factors but to a lesser extent. Exposure time relates to the duration over which organisms or cells are subjected to nanoparticles, affecting the overall toxic impact. Similarly, the core size of nanoparticles can influence their interaction with biological systems, affecting how they are absorbed, distributed, and metabolized.

This Decision Tree model's results demonstrate the multifaceted nature of nanoparticle toxicity, where multiple variables, including chemical composition, physical properties, and usage conditions, collectively determine their safety profile. Such insights are crucial for developing safer nanomaterials and for regulatory purposes, ensuring that all significant factors are considered in toxicity assessments.

The line chart (fig. 11) depicting the impact of various factors on nanoparticle toxicity presents a clear hierarchy of influence among the variables studied. At the forefront, dosage stands out as the most significant factor affecting toxicity, suggesting that the quantity of nanoparticles administered plays a pivotal role in determining their toxic potential. This aligns with the principle that greater amounts of a substance can lead to more pronounced biological effects, including potential harm.

Following dosage, surface area is the next most influential factor. The extent of a nanoparticle's surface area affects its interactions with biological environments, as a larger surface area provides more contact points for biological reactions, potentially increasing toxicity.Exposure time also plays a crucial role, with longer durations of exposure increasing the likelihood of toxic effects as the nanoparticles have more time to interact with biological systems.The number of oxygen atoms (NOxygen) and core size are next in terms of impact. NOxygen's influence highlights the importance of chemical composition in toxicity, while core size's effect underscores the role of physical dimensions in how nanoparticles behave in biological systems.Lastly, hydrosize, though still significant, has the least impact among the factors listed. This smaller influence suggests that while the hydrodynamic size of nanoparticles affects their biological interactions, it may be less critical compared to other factors such as chemical makeup or physical structure.This analysis from the line chart provides valuable insights for understanding how various properties of nanoparticles contribute to their toxicity, guiding safer nanoparticle design and usage.

The bar chart (fig. 12) provides a visual representation of the correlation between various physicochemical properties of nanoparticles and their toxicity. Each bar in the chart corresponds to a different property, showing whether its relationship with toxicity is direct (positive correlation) or indirect (negative correlation). A direct effect, indicated by bars extending to the right, signifies that an increase in the property enhances toxicity, suggesting that these characteristics may need to be controlled or minimized in safe nanoparticle design. Conversely, bars extending to the left indicate an indirect (negative) effect, where higher values of the property are associated with reduced toxicity, highlighting these as potentially protective or mitigating factors in the context of nanoparticle toxicity. This visualization aids in quickly identifying which properties are most influential in contributing to or detracting from the toxicological impacts of nanoparticles.

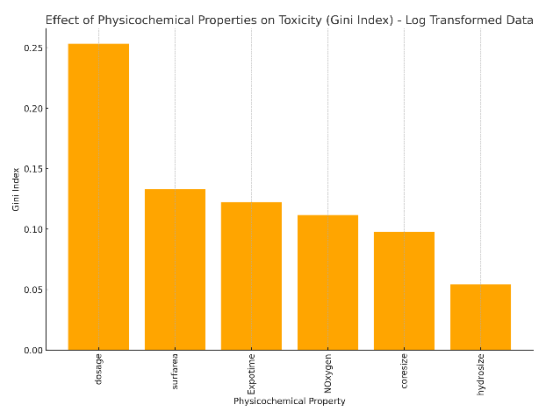

*Figure 11*

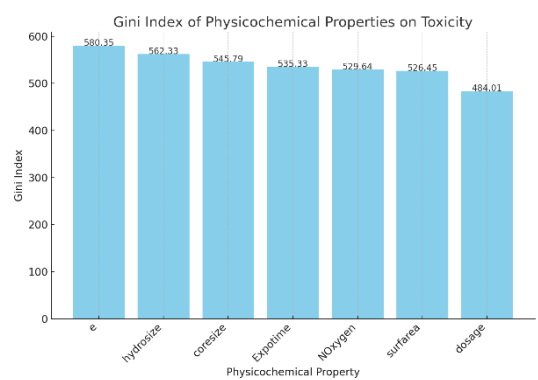

*Figure 12*

## Conclusion:

In conclusion, the research successfully developed and tested various machine learning models to predict the toxicity of nanoparticles without relying on traditional in vivo or in vitro experiments. By utilizing computational tools, particularly machine learning, the study was able to forecast toxicity based on key physicochemical properties of nanoparticles, such as particle size, surface charge, and exposure conditions. The models used included Decision Trees, Random Forest, XGBoost, Logistic Regression, and k-Nearest Neighbors, with XGBoost demonstrating the highest accuracy and precision. The development of these predictive models offers a more cost-effective, faster, and less ethically controversial method for assessing the safety of nanoparticles, reducing the need for extensive animal testing.

Furthermore, the research aimed to create a graphical user interface (GUI) that enables users to input nanoparticle characteristics and receive toxicity predictions in real-time. This practical application of the research allows for greater accessibility and utilization of the models by non-experts in the field, potentially facilitating safer nanoparticle development across industries.This innovative approach demonstrates the effectiveness of computational techniques in addressing complex problems in nanotoxicology, marking a significant advancement in reducing the dependency on experimental testing while maintaining high accuracy in predictions.